 \documentclass{article}
\usepackage{cite}
\usepackage{amsmath,amssymb,amsfonts}
\usepackage{algorithmic}
\usepackage{graphicx}
\usepackage{textcomp}

\newcommand{\bqn}{\begin{eqnarray}}
\newcommand{\eqn}{\end{eqnarray}}
\newcommand{\bq}{\begin{eqnarray*}}
\newcommand{\eq}{\end{eqnarray*}}

 \newcommand{\bb}[1]{{\mathbf{#1}}}
 
\usepackage{multirow}
\usepackage{adjustbox}

\usepackage{url}
\usepackage{hyperref}
\hypersetup{colorlinks,%
citecolor=black,%
filecolor=black,%
linkcolor=red,%
urlcolor=black}%

\begin{document}

\title{Statistical Analysis of Dynamic \\
Functional Brain Networks in Twins}
\author{Moo K. Chung, Shih-Gu Huang, Tananun Songdechakraiwut,\\ 
 Ian C. Carroll, H. Hill Goldsmith
\thanks{
This study was supported by NIH research grants EB022856, EB028753, P30HD003352, U54HD09025 and UL1TR002373.
(Corresponding author: Moo K. Chung. e-mail: mkchung@wisc.edu)}
\thanks{
M.K. Chung, S.-G. Huang  and T. Songdechakraiwut are with the Department of Biostatistics and Medical Informatics, University of Wisconsin, Madison, WI, USA. Chung is also with Waisman Laboratory for Brain Imaging and Behavior, University of Wisconsin, Madison, WI 53706, USA.}
\thanks{I. C. Carroll and H. H. Goldsmith are with the Waisman Center \& Department of Psychology, University of Wisconsin, Madison, WI, USA.}
}

\maketitle

\begin{abstract}
Recent studies have shown that functional brain brainwork is dynamic even during rest. 
A common approach to modeling the brain network in whole brain resting-state fMRI is to compute the correlation between anatomical regions via sliding windows. However, the direct use of the sample correlation matrices is not reliable due to the image acquisition, processing noises and the use of discrete windows that often introduce spurious high-frequency fluctuations and the zig-zag pattern in the estimated time-varying correlation measures. To address the problem and obtain more robust correlation estimates,  we propose the heat kernel based dynamic correlations. We demonstrate that the  proposed heat kernel method can smooth out the unwanted high-frequency fluctuations in correlation estimations and achieve higher accuracy in identifying dynamically changing distinct states. The method is further used in determining if such dynamic state change is genetically heritable using a large-scale twin study. Various methodological challenges for analyzing paired twin dynamic networks are addressed. 
\end{abstract}


\section{Introduction}

Findings of resting-state fMRI have revealed synchrony between spontaneous  blood-oxygen-level-dependent (BOLD) signal fluctuations in sets of distributed brain regions despite the absence of any explicit tasks \cite{Chang2010,Hutchison2013,Hutchison2015,Preti2017}. The time-invariant static measures of functional connectivity are often computed over the entire scan duration. However, this oversimplification reduces the complex dynamics of the resting-state functional connectivity to the time average. Recent studies have suggested the dynamic changes in functional connectivity over time even during rest, referred to as the \textit{dynamic functional connectivity}, can be a meaningful biomarker \cite{Chang2010,Hutchison2013,Hutchison2015,Preti2017}.

The most common approach to modeling dynamic connectivity is through the sliding windows (SW), where correlations between brain regions are computed over the windows \cite{keilholz2013dynamic,kucyi2014dynamic,Allen2014,Hutchison2013,Hutchison2015,Shakil2016,Hindriks2016,mokhtari2019sliding,Zalesky2015}. Various SW methods have been proposed including  the tapered sliding window (TSW), which uses  a  square window convolved with a Gaussian kernel \cite{Allen2014,lindquist2014evaluating,abrol2017replicability,mokhtari.2019},
Hamming window \cite{handwerker2012periodic}, Tukey window \cite{Rashid2014} and exponentially decaying window \cite{lindquist2014evaluating}. However, the sidelobes of the discrete window functions in the spectral domain  cause high-frequency fluctuations in the dynamic correlations in all these methods \cite{oppenheim2001discrete}. Further, correlation computation within windows is very sensitive to outliers \cite{devlin.1975}.  To address these problems, we propose  the heat kernel method, which computes the dynamic correlations without the heat kernel \cite{huang.2019.DSW}. We show that the proposed heat kernel method significantly reduces the unwanted high-frequency noises in the estimation of dynamic correlations. One can summarize the whole-brain dynamically changing functional connectivity into a smaller set of \textit{dynamic connectivity states}, defined as distinct transient connectivity patterns that repetitively occur throughout the resting-state scan \cite{Hutchison2013,Hutchison2015}. They are reliably observed across different subjects, groups and sessions \cite{Yang2014,Choe2017, Allen2014,Damaraju2014,Rashid2014, Barttfeld2015,Hutchison2015,
Rashid2016,Marusak2017}. We show that the proposed heat kernel method is more robust than SW-  and TSW-methods in identifying and discriminating states in  a large-scale twin imaging study. 

There are other methods such as instantaneous phase synchrony analysis (IPSA) \cite{le.2001, guevara.2005,pedersen.2017, pedersen.2018} or phase angle spatial embedding (PhASE) \cite{morrissey.2018} or weighted phase lag index (WPLI) \cite{xing.2019}.  These methods embed data into the complex plane and measure the synchrony between the phase angles. These methods were not considered in this paper. Time varying sparse network models such as graphical-LASSO \cite{cai.2018} and sparse hidden Markov models \cite{zhang.2019} are also not considered in this paper as well.

Brain functions are heavily influenced by genetic factors.  Twin brain imaging studies offer valuable genetic information, based on the differing genetic similarity of the two zygosities, which allows estimation of genetic effects. Monozygotic (MZ) twins share 100\% of genes while  dizygotic (DZ) twins share 50\% of genes \cite{falconer.1995,chung.2017.IPMI}. MZ-twins are more similar or concordant than DZ-twins for cognitive aging, cognitive dysfunction, and Alzheimer's disease \cite{reynolds.2015}. The difference between MZ- and DZ-twins directly quantifies the extent to which imaging phenotypes  are influenced by genetic factors. If MZ-twins show more similarity on a given feature compared to DZ-twins, this provides evidence that genes significantly influence that feature. Previous twin brain imaging studies mainly used univariate imaging phenotypes such as cortical surface thickness \cite{mckay.2014}, fractional anisotropy \cite{chiang.2011}, or functional activation \cite{blokland.2011,glahn.2010,smit.2008} in  determining heritability in the regions of interest \cite{jansen.2015,richmond.2016}.  Measures of network topology and features are worth investigating as multivariate phenotypes \cite{bullmore.2009}. However, many existing twin brain network studies mainly focus on determining the genetic contributions of {\em static} networks  \cite{glahn.2010,blokland.2011,fornito.2011,chung.2019.NN}.    

The main purpose of  this paper is to demonstrate that genes also shape the overall dynamic brain networks even during rest. 
This goal is achieved based on two main  contributions against existing literature and our previous study \cite{huang.2019.DSW}: 

1) We present a novel heat kernel method for computing dynamic correlation network that improves the estimation performance over the use of discrete windows with detailed mathematical and numerical justifications.

2) The proposed method is applied to 232 twins (130 MZ-twins and 102 DZ-twins) to determine the heritability of such state changes. This requires addressing various methodical issues that are specific to the paired network setting, which is rarely encountered in non twin imaging studies.

\section{Methods}

\begin{figure}[t]
\centering
\includegraphics[width=1\linewidth]{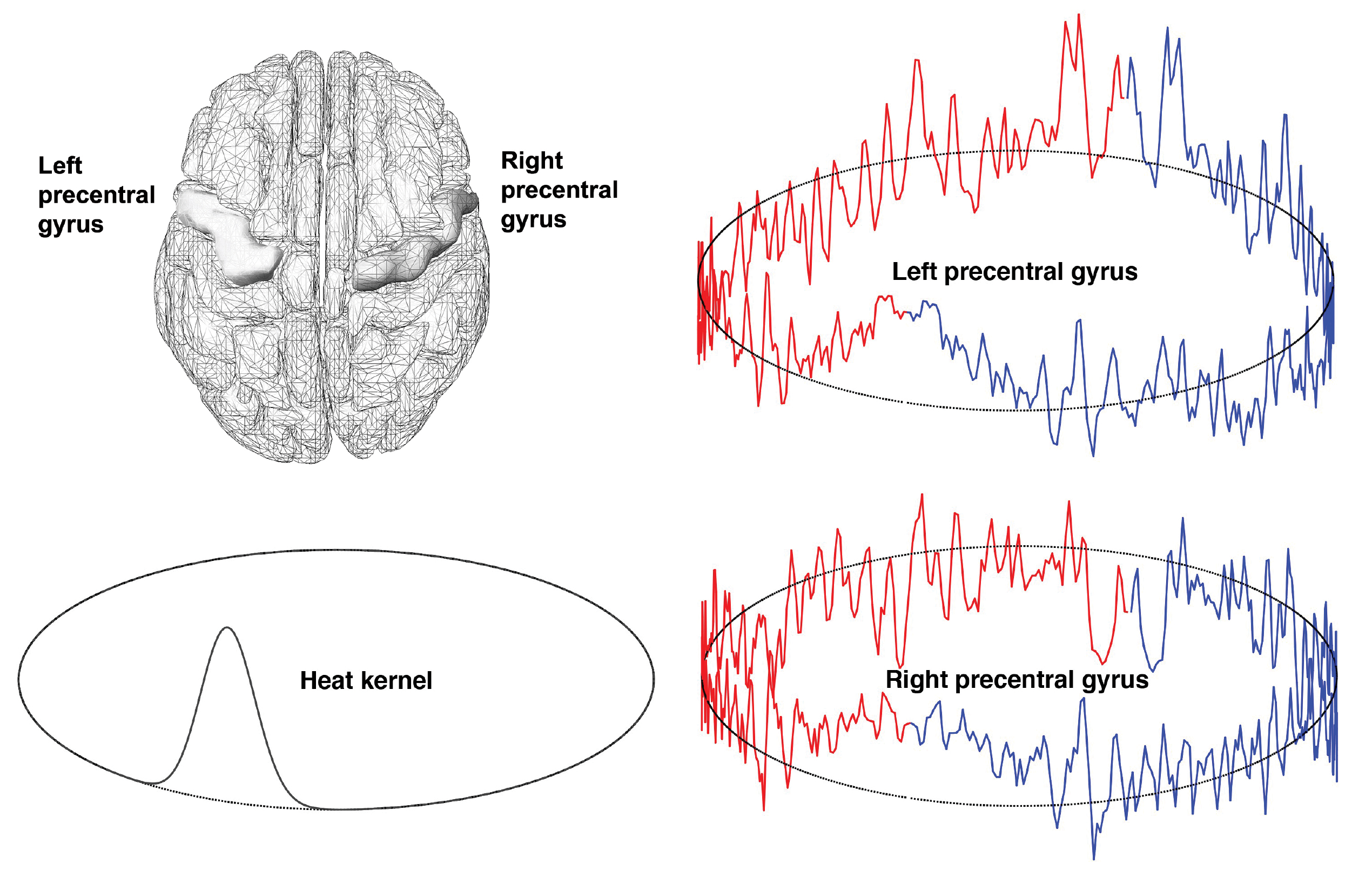}
\caption{Top left:  brain regions using AAL template. The left  and right  precentral gyri are marked. 
Right column: average fMRI signals within left  and right  precentral gyri (blue) projected onto a circle and connected with their mirror reflections (red). Bottom left: heat kernels $K_s(t,t')$ plotted at $t=0$.}
\label{fig:circle}
\end{figure}

\subsection{Correlation over sliding windows} 
Widely used windowed dynamic correlations can be formalized as follows. Consider  time series 
\bq \bb x &=& (x_0,x_1, \cdots, x_{T-1})\\
\bb y &=& (y_0,y_1,..,y_{T-1})\eq
with $T$ time points. To reduce the boundary effect in 
windowed methods \cite{jones1993simple}, we connect the data at the end time points $0$ and $T-1$ by mirror reflection and  make them into the circular data with $2T$ data points:
\bq \bb x\!\! &=& \!\!(\cdots, x_2, x_1, x_0, x_1, x_2, \cdots, x_{T-1}, x_{T-1}, x_{T-2}, \cdots),\\
\bb y\!\! &=& \!\!(\cdots, y_2, y_1, y_0, y_1, y_2, \cdots, y_{T-1}, y_{T-1}, y_{T-2}, \cdots).\eq
Figure \ref{fig:circle} displays the average fMRI in the left and right precentral gyri connected at the first ($t=0$) and the 295-th scan ($t=1$) in the circular fashion.

Let  $W_i = [ \lfloor i-\frac{m}{2}+1  \rfloor,  \lfloor i+\frac{m}{2} \rfloor ]$ be the window of size $m$ centered at time point $i$, where $\lfloor \; \rfloor$ is the floor function.
The mean and variance of $\bb x$  within window $W_i$ are computed as
\bq \overline{\bf x}_i &=&  \sum_{j \in W_i}  w_jx_j, \\
{\sigma_{\bf x}^2}_i &=& \sum_{j \in W_i} w_j(x_j- \overline{\bf x}_i)^2,
\eq
where $w_j$ are the weights following the shape of the window  and satisfies
$$\sum_{j \in W_i} w_j =1.$$
For the usual square or rectangular window, we have $w_j = 1/m$. In \cite{Allen2014}, the tapered window, which is the convolution of the square window with a Gaussian kernel, is used so that the data points will gradually enter and exit when moving across time \cite{lindquist2014evaluating}. Whether the window is square  or tapered square, correlation is given by

\bqn \rho_i=\frac{\sum_{j \in W_i}  w_{j} (x_j- \overline{\bf x}_i)(y_j- \overline{\bf y}_i)}
{{\sigma_{\bf x}}_i {\sigma_{\bf y}}_i}. \label{eq:discreterho}
\eqn
Then the sliding window $W_i$ at time point $i$ will slide one time point a time. Due to the reflection of data,  for a time series with $T$ time points, we will have $2T$ overlapping sliding windows.

\begin{figure}[t]
\centering
\includegraphics[width=1\linewidth]{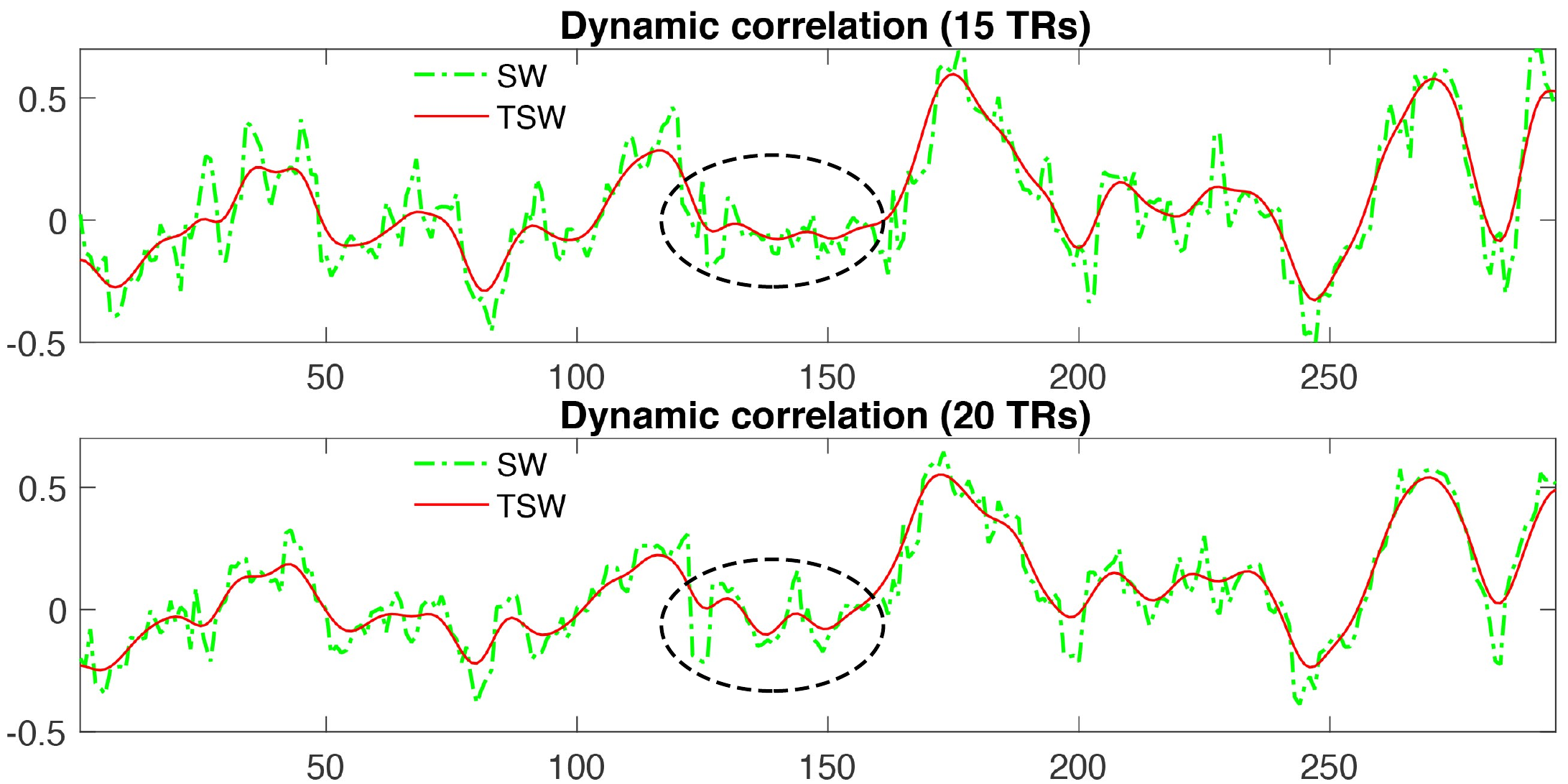}
\caption{Dynamic correlations computed by the SW- and TSW-methods with window size measured as the full width at half maximum (FWHM)  15 (top) and 20 (bottom) TRs. The SW-method shows the severe zig-zag pattern. The TSW-method reduced the zig-zag pattern but we can still observe high frequency fluctuations.
These zig-zag pattern and high-frequency fluctuations were not eliminated and  became even worse in some  intervals (dashed circles) when  larger window size was used indicating they are in fact artifacts produced by the use of discrete windows.}
\label{fig:SquareWinCorr}
\end{figure}

\begin{figure}[t]
\centering
\includegraphics[width=1\linewidth]{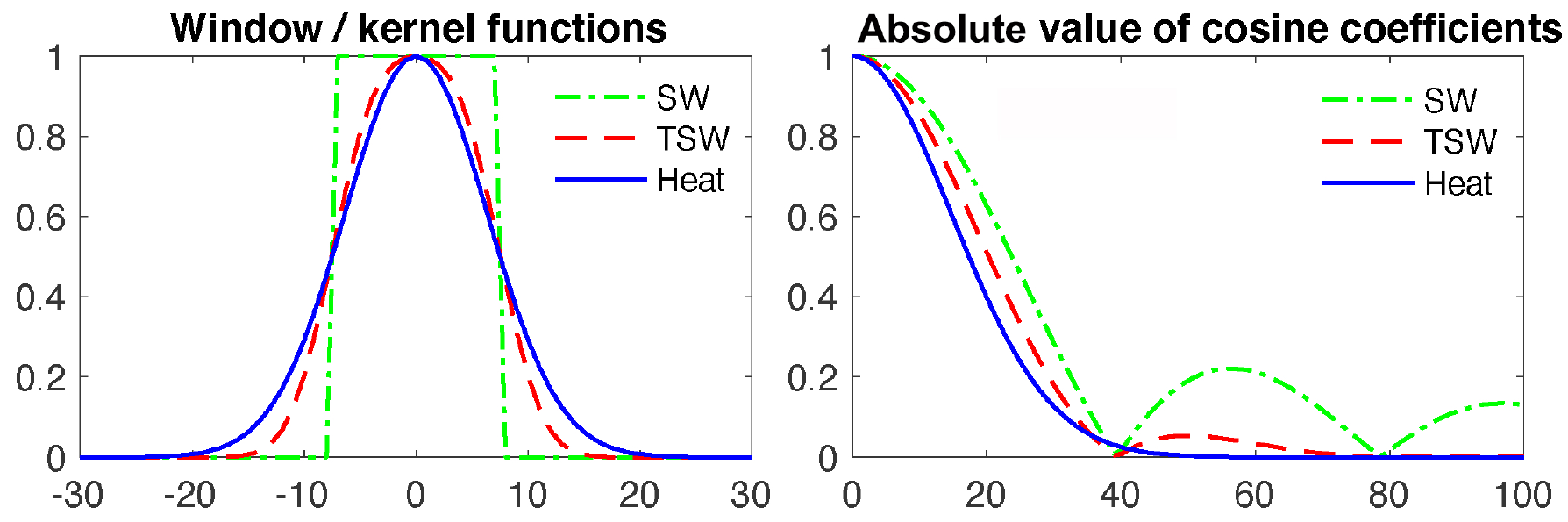}
\caption{Left: The square window, tapered window and heat kernel with the same FWHM equivalent to heat kernel bandwidth $s=2.3\times 10^{-4}$ The heat kernel is defined on a circle continuously without endpoint or boundary. Here, we only showed the kernel at 60 time points. Right: The absolute values of the first 100 cosine series expansion coefficients of the widow and kernel functions. The sidelobes  of the window functions in spectral domain will cause severe fluctuations that caused severe zig-zag pattern.}
\label{fig:sidelobes}
\end{figure}

Figure \ref{fig:SquareWinCorr} displays an example of the SW-method with window sizes 15 and 20 TRs. 
The SW-method suffers from  severe zig-zag patterns caused by the use of the discrete window, which could not be effectively reduced even if we increase the window size from 15 to 20 TRs. Figure \ref{fig:SquareWinCorr} also displays the TSW-method using the square window of sizes 15 and 20 TRs convolved with the Gaussian kernel with bandwidth 3 TRs \cite{lindquist2014evaluating}. The TSW-method reduces the zig-zag pattern in SW-method significantly, but the TSW-method still shows rapid high frequency fluctuations. Even more high-frequency fluctuations occur in some time intervals when a larger window is used. This indicates that these fluctuations are in fact artifacts produced by the use of discrete windows. Such zig-zag patterns and high-frequency fluctuations in the SW- and TSW-methods are caused by the sidelobes of the window functions in the spectral domain \cite{oppenheim2001discrete} (Figure~\ref{fig:sidelobes}). To address the problem caused by the use of discrete window, we propose to use the {\em heat kernel} defined over all time points.

\subsection{Heat kernel convolution on a circle}

We first define the heat kernel on the unit interval and extend it to a circle. 
The heat kernel has been mainly used on nonstandard geometry such as the brain cortical surface \cite{chung.2007.TMI}. The heat kernel can be easily constructed on the circle as well. Consider 1D heat diffusion of functional data $f(t)$ on unit interval $[0,1]$:
\bqn \frac{\partial}{\partial {s}} h(t,s) = \frac{\partial^2}{\partial t^2} h(t,s) \label{eq:diffusion1} \eqn
at diffusion time $s$ with initial condition $h(t,{s}=0)=f(t)$. Note the initial functional data do not have to be smooth or differentiable. Then the unique solution  is given by  the weighted Fourier series (WFS) \cite{chung.2007.TMI,chung.2010.SII}
\bqn h(t,{s})=\sum_{l=0}^\infty e^{- l^2\pi^2 {s}}c_{fl}\psi_l(t), 
\label{eq:hts01} 
\eqn
where $\psi_0(t) =1$, $\psi_l(t) = \sqrt{2} \cos ( l \pi t)$ are the cosine basis 
and  $c_{fl}$ are the expansion coefficients of $f$ with respect to $\psi_{l}$:
$$ c_{fl}=\int_{0}^1f(t)\psi_l(t)dt.$$ 
We can rewrite (\ref{eq:hts01}) as convolution
\bq h(t,{s})=\int_{0}^1K_s(t,t')f(t')dt', \label{eq:hts}\eq
where heat kernel $K_s(t,t')$ is defined as
\bq
K_s(t,t')=\sum_{l=0}^\infty e^{- l^2\pi^2 {s}}\psi_l(t)\psi_l(t').
\label{eq:heat}
\eq
The heat kernel is a probability distribution satisfying 
$$\int_{0}^1 K_s(t,t')dt' = 1.$$
The diffusion time $s$, also referred to as the kernel bandwidth,  controls the amount of diffusion.

\begin{figure}[t]
\includegraphics[width=1\linewidth]{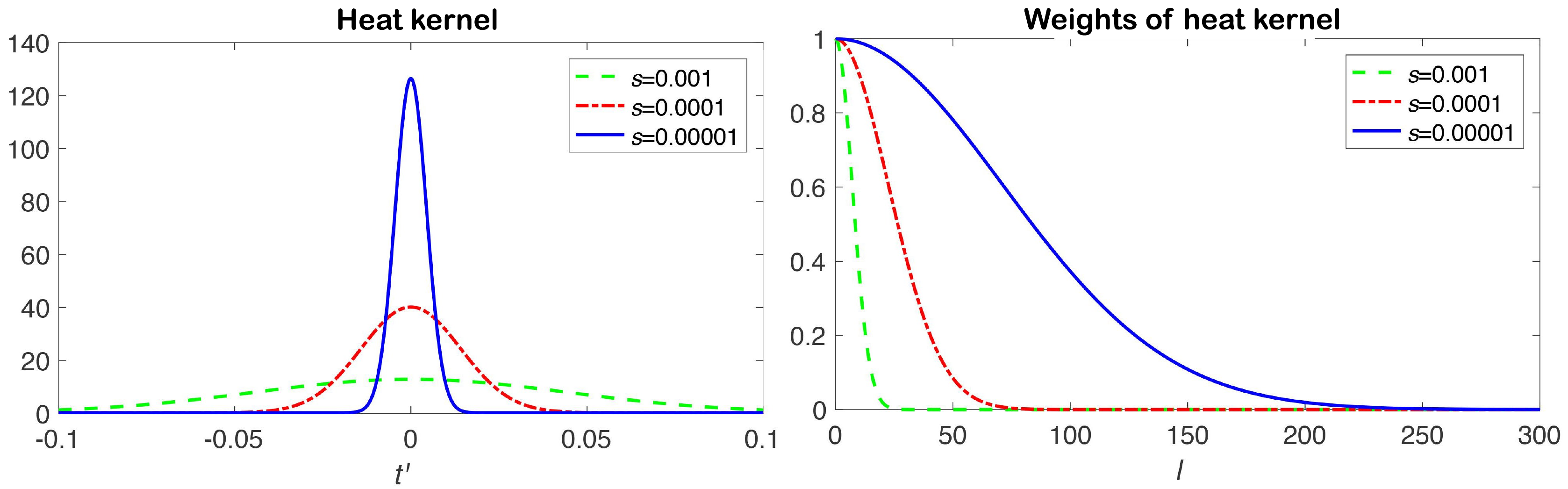}
\caption{Left: heat kernels $K_s(t,t')$  at $t=0$  with different diffusion time or bandwidth $s$.
The heat kernel has larger FWHM when $s$ increases.
Right:  weights $e^{- l^2\pi^2 s}$ of the heat kernels. As $s$ increases, the weights in the high frequencies become smaller compared to low frequencies, and more high-frequency components will be smoothed out.}
\label{fig:Heat}
\end{figure}

Now we project $f(t)$ defined on $[0,1]$ onto  the circle $\mathcal{C}$ with circumference 2 by the mirror reflection in the following way
$$g(t) = f(t) \mbox{ if } t \in [0,1], \quad g(t)=f(2-t) \mbox{ if } t \in [1,2].$$ 
Then we solve diffusion equation (\ref{eq:diffusion1}) with initial condition $h(t,s=0) = g(t)$, which is periodic on $\mathcal{C}$. 
It can be shown that the solution is given by
\bq h(t,{s})=\int_{0}^2 K_s(t,t')g(t')dt'. \label{eq:hts}\eq
Using the mirror symmetry, we extend the domain of  cosine basis $\psi_l$ and heat kernel 
\bq \psi_l(t) &=& \psi_l(2-t)\\
 K_s(t,t) &=& K_s(t,2-t)\eq
 for $t \in [1,2]$. Then we have
\bq
\int_{0}^2 K_s(t,t')dt' = \int_{0}^1 K_s(t,t')dt' + \int_{0}^1 K_s(t,2-t')dt' = 2.
\eq
Unlike SW and TSW window functions, there is no endpoint or boundary in the heat kernel defined on a circle with non-zero values over the entire circle . On the circle, which is a curved manifold, heat kernel has a thicker tail compared to truncated Gaussian kernel.  As bandwidth ${s}$ increases, the tail regions  get thicker and eventually we have \cite{chung.2007.TMI}
$\lim_{s \to \infty} K_s(t,t') = 1.$
Figure \ref{fig:Heat} plots the heat kernels $K_s(t,t')$ at fixed $t=0$ and the weights $e^{- l^2\pi^2 s}$ 
for different bandwidth $s$. The heat kernel does not have the  sidelobe problem  as in the discrete window functions. 

Similarly, we also have
\begin{align*}  &\int_{0}^2 K_s(t,t')g(t')dt' \\
&= \int_{0}^1 K_s(t,t')g(t')dt' + \int_{0}^1 K_s(t,2-t')g(2-t')dt'\\
&=  2 \int_{0}^1 K_s(t,t')g(t')dt'.
\end{align*}
Hence, heat kernel smoothing on the circle can be simply done by smoothing in unit interval $[0, 1]$.

In the numerical implementation,  the cosine series coefficients $c_{fl}$ are  estimated using the least squares method \cite{chung.2010.SII}.
We set the expansion degree to equate the number of time points, which is 295. Window size of 10 to 20 TRs were used in SW and TSW methods. This is equivalent to heat kernel bandwidth of $s=10^{-5}$ and  $10^{-4}$ in terms of FWHM.

 \subsection{Heat kernel based dynamic correlation}
The heat kernel based dynamic correlation between $x(t)$ and $y(t)$ over $[0,1]$ is
\bqn  \rho (t) = \frac{\int_{0}^1 K_s(t,t')x(t')y(t')dt' -\mu_x(t)\mu_y(t)}
{\sigma_{x}(t) \sigma_{y}(t) }, \label{eq:rxy} \eqn
where 
\bq \mu_x(t)&=&\int_{0}^1 K_s(t,t')x(t')dt',\\
\sigma_{x}^2(t)&=&\int_{0}^1 K_s(t,t')x^2(t')dt'-\mu^2_x(t)\eq
are the  dynamic mean and variance of $x(t)$. $\mu_y(t)$ and $\sigma_{y}^2(t)$ are defined similarly. 
The heat kernel based correlation generalizes the integral correlations with the additional weighting term \cite{huang.2019.ISBI} and 
also generalizes the discrete windowed correlation (\ref{eq:discreterho}). 

\begin{figure}[t]
\centering
\includegraphics[width=1\linewidth]{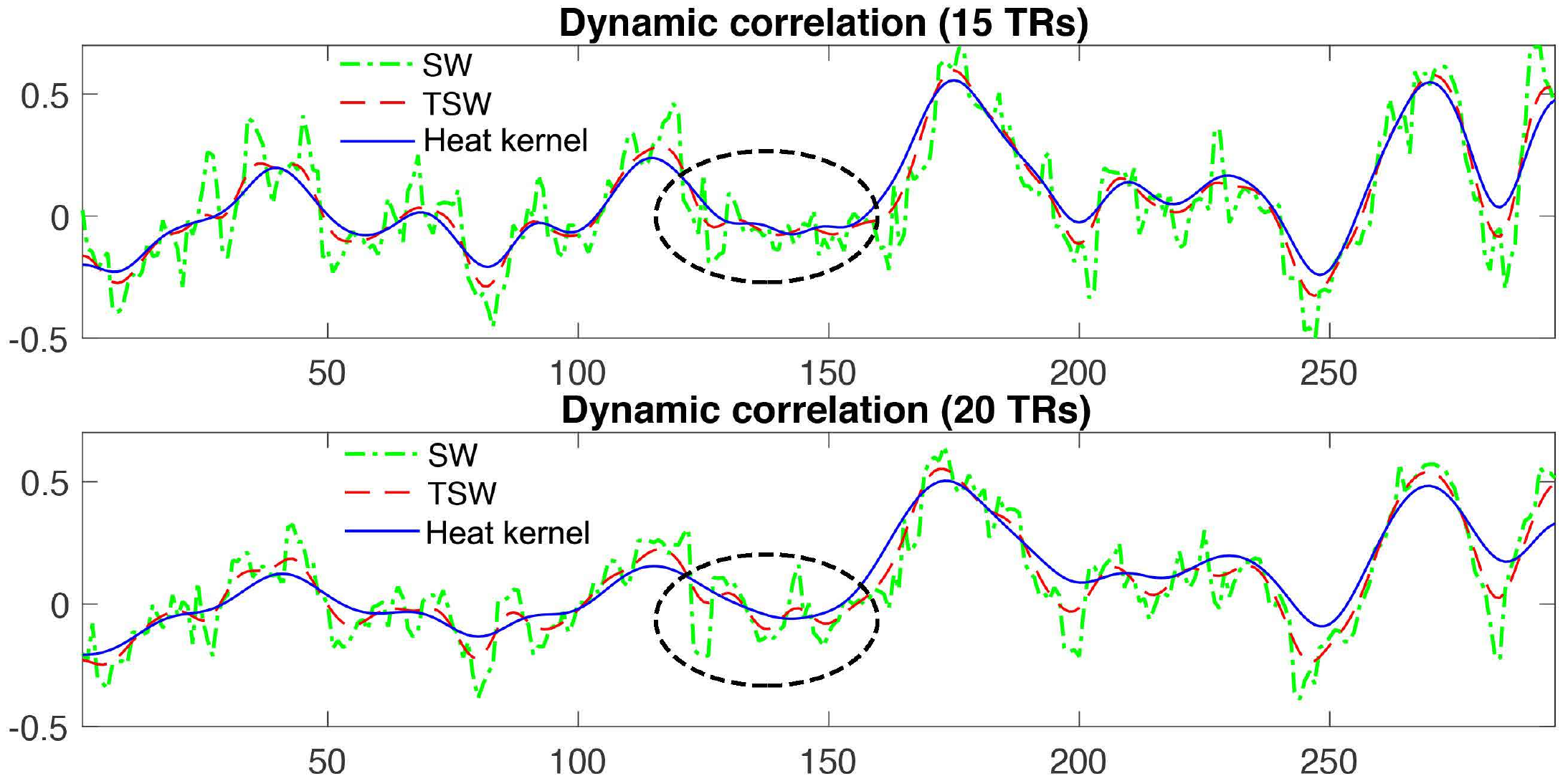}
\caption{The SW-, TSW- and proposed heat kernel methods with FWHM 15 (top) and 20 (bottom) TRs. The heat kernel method eliminated most of the zig-zag pattern and high-frequency fluctuations in the SW- and TSW-methods.
}
\label{fig:HeatCorr}
\end{figure}

Suppose we further represent  functional data $x(t)$ and $y(t)$ using the  cosine basis as \cite{chung.2010.SII}
$$x(t)=\sum_{l=0}^\infty c_{xl}\psi_l(t), \ \ \ y(t)=\sum_{l=0}^\infty c_{yl}\psi_l(t),$$
where $$c_{xl}=\int_{0}^1x(t)\psi_l(t)dt, \quad c_{yl}=\int_{0}^1y(t)\psi_l(t)dt$$ are the cosine series coefficients. 
Similarly we expand $x(t)y(t)$, $x^2(t)$ and $y^2(t)$ using the cosine basis and obtain coefficients $c_{xyl}$, $c_{xxl}$ and $c_{yyl}$. Then heat kernel based  dynamic correlation (\ref{eq:rxy}) can be further written as

\begin{align} \label{eq:windowless}
\rho (t) =\frac{\sum_{l=0}^\infty e^{-l^2 \pi^2 s}  c_{xyl}\psi_l(t) -\mu_x(t)\mu_y(t)}
{\sigma_{x}(t) \sigma_{y}(t) },
\end{align}
with
\bq \mu_x(t) &=& \sum_{l=0}^\infty e^{-l^2 \pi^2 s} c_{xl}\psi_l(t),\\
\sigma_{x}^2(t) &=&\sum_{l=0}^\infty e^{-l^2 \pi^2 s}c_{xxl}\psi_l(t)-\mu^2_x(t).\eq
Correlation (\ref{eq:windowless}) is the formula we used to compute the dynamic correlation in this study.

In SW- and TSW-methods, smaller windows can capture more short-lived variations than larger windows \cite{Shakil2016}, but will increase the risk of creating high-amplitude variations and spurious fluctuations even  when the brain connectivity is actually static \cite{leonardi2015spurious,lindquist2014evaluating,mokhtari2019sliding}.
In this study, following \cite{Allen2014,lindquist2014evaluating}, square windows of size 15  and 20 TRs (i.e., 30 and 40 seconds) were used in the SW-method. For TSW-method, the tapered windows were obtained by convolving the square windows   with a Gaussian kernel with  bandwidth 3 TRs. For the proposed heat kernel method, bandwidth $s=2.3\times 10^{-4}$ and $4.1\times 10^{-4}$, which give equivalent FWHM 15 and 20 TRs respectively. Figure \ref{fig:HeatCorr} displays the heat kernel based dynamic correlation. The proposed heat kernel method eliminated most of the zig-zag pattern and high-frequency fluctuations in the SW- and TSW-methods.

\begin{figure}[t]
\centering
\includegraphics[width=1\linewidth]{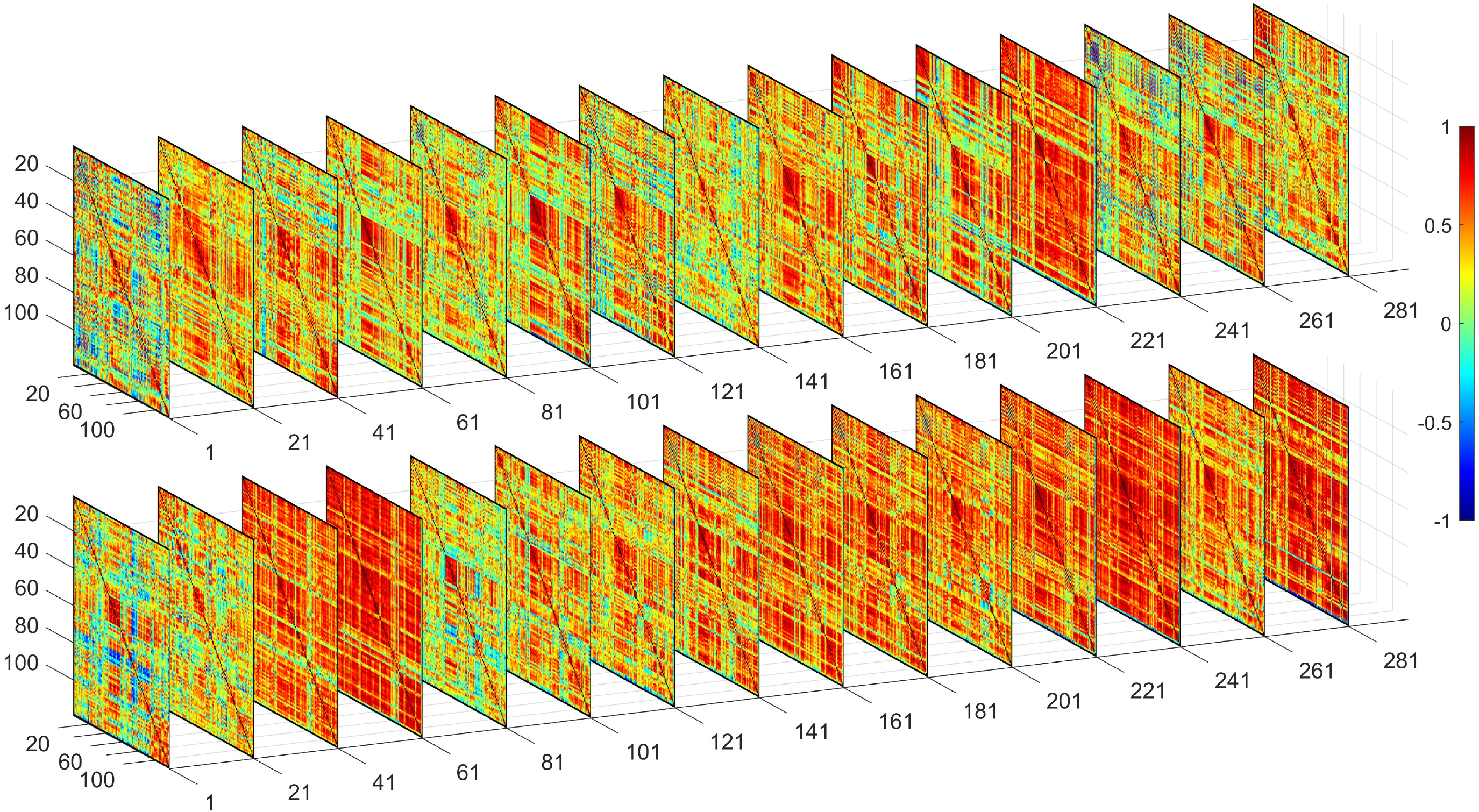}
\caption{$116\times116$  dynamic correlation matrices of a MZ-twin at time points $1,21,41,...,281$  using the proposed heat kernel method with FWHM 15 TRs. In the actual analyses, we used all the time points. 
}
\label{fig:AAL_DynCorrMtx}
\end{figure}

\subsection{Estimation of distinct state space}\label{subsec:states}

For $p$ brain regions, we estimate $p\times p$ dynamically changing correlation matrices $C_i(t)$  for the $i$-th subject at time points $t=t_1,...,t_T$. Let $\mathbf{d}_{ij}$ denote the vectorization of  the upper triangle of $p \times p$ matrix $C_i(t_j)$. Figure \ref{fig:AAL_DynCorrMtx} displays the dynamically changing correlation matrice of a single MZ-twin is shown, where the proposed heat kernel method with FWHM 15 TRs was used. The collection of $\mathbf{d}_{ij}$ over $T$=295 time points  and $n=$479 subjects is then feed into the $k$-means clustering  in identifying the recurring brain connectivity states that are common across subjects  at the group level \cite{Allen2014,Barber2018}. For each subject, the  state visits at  295 time points are then represented as a time series of integers between 1 and $k$. These discrete states serve as the basis of investigating  the dynamic pattern brain connectivity \cite{choe2017comparing,Barber2018,ting2018estimating}.

\begin{figure}[t]
\centering
\includegraphics[width=1\linewidth]{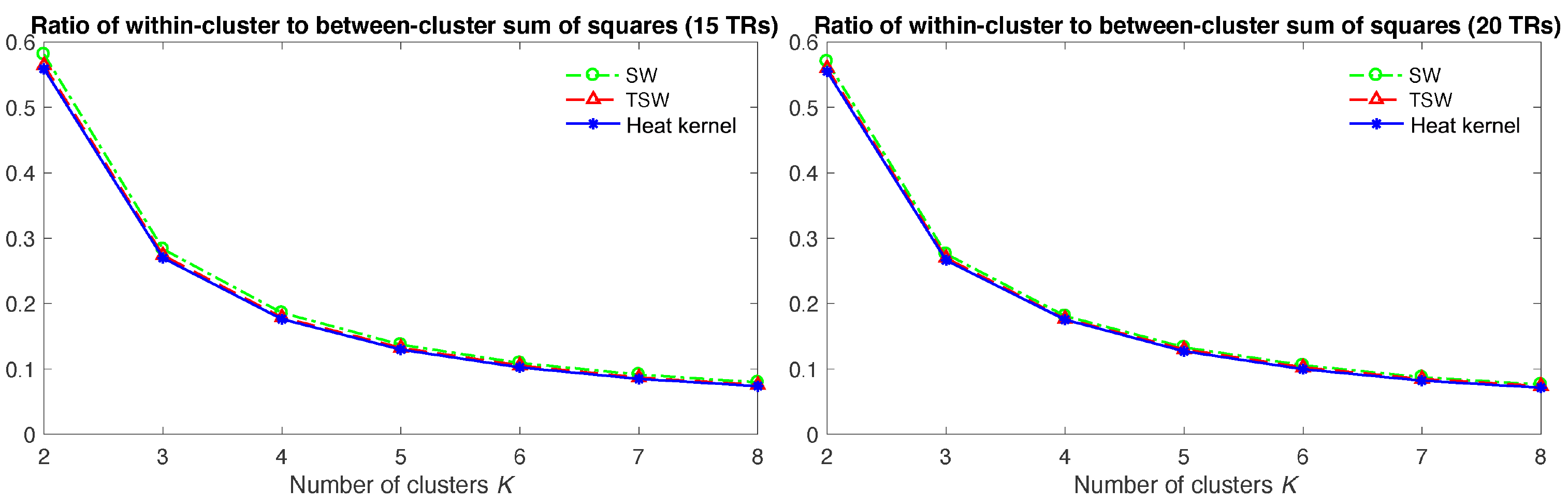}
\caption{The ratio of within-cluster to between-cluster sum of squared distances versus the number of clusters $K=2,...,8$ for window/kernel FWHM 15 (left) and 20 (right) TRs.
By  the elbow method, three clusters were chosen since the slope changes the most drastically from steep to shallow  at the elbow point $K=3$.}
\label{fig:elbow}
\end{figure}

\begin{figure}[t]
\centering
\includegraphics[width=1\linewidth]{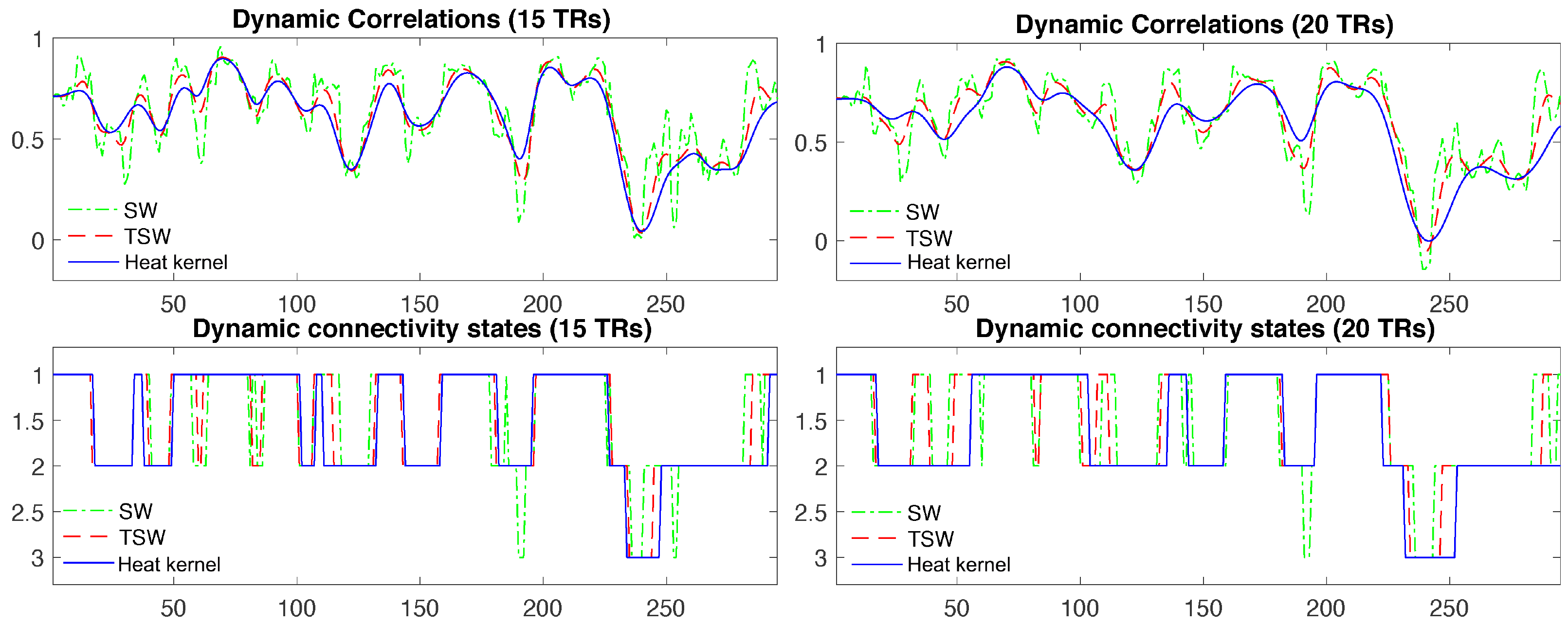}
\caption{Dynamic correlation between the left and right precentral gyri (top) and estimated state space (bottom). 
The SW-, TSW- and proposed heat kernel methods with FWHM 15 (left) and 20 (right) TRs were used.
The heat kernel method estimated the dynamic correlations more smoothly over time with the least number of state changes.}
\label{fig:interhemisphere}
\end{figure}

For the convergence of $k$-means clustering, the clustering was repeated 100 times with different initial centroids and  the best result with the lowest sum of squared distances was chosen. Simply running the $k$-means clustering algorithm with a single fixed initial centroids usually do not guarantee the convergence of $k$-means clustering. The optimal number of cluster $k$ was determined by the elbow method \cite{Allen2014,Rashid2014,nomi2016dynamic,abrol2017replicability,choe2017comparing,lehmann2017assessing,Barber2018,ting2018estimating}. For each value of $k$, we computed the within-cluster  and between-cluster sums of squared distances. By the elbow method, we chose $k=3$ which gives the largest slope change in the ratio of within-cluster to between-cluster sum of squares (Figure \ref{fig:elbow}). Figure \ref{fig:interhemisphere} displays
the result of $k$-means clustering on the heat kernel method. 

The dynamics of state change can be modeled as a Markov chain \cite{gilks1995markov,baker2014fast}.
For subject $i$, let $s_{i}(t)\in\{1,2,3\}$ be the state label at time $t$. The transition probability of moving from state $k_1$ to state $k_2$, i.e.,
$$P\left(s_{i}(t)=k_2|\ s_{i}(t-1)=k_1\right)$$ 
is then used to quantify the dynamics of state changes. We also used the occupancy rate of state $k_j$ given by
$$ \frac{1}{nT} \sum_{i=1}^n\sum_{t=1}^{T} \mathcal{I}(s_{i}(t) = k_j ),$$
where $\mathcal{I}$ is the indicator function, $n=479$ subjects and $T=295$ time points.

\begin{figure}[t]
\centering
\includegraphics[width=1\linewidth]{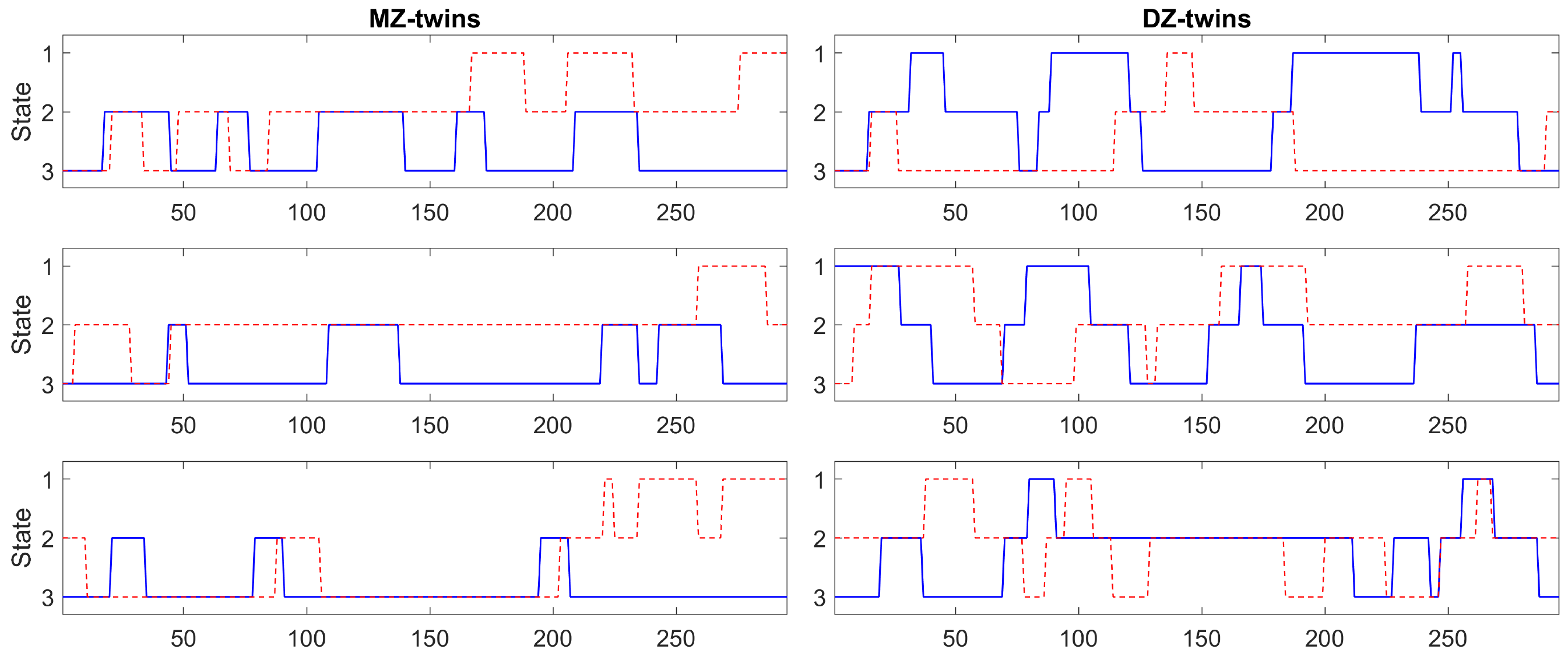}
\caption{State visits for 3 MZ-twins (left) and 3 DZ-twins (right). They are selected randomly. We are interested in determining the heritability such state visits.}
\label{fig:statevisits}
\end{figure}

\subsection{Heritability estimate in twins}

\begin{figure}[t]
\includegraphics[width=1\linewidth]{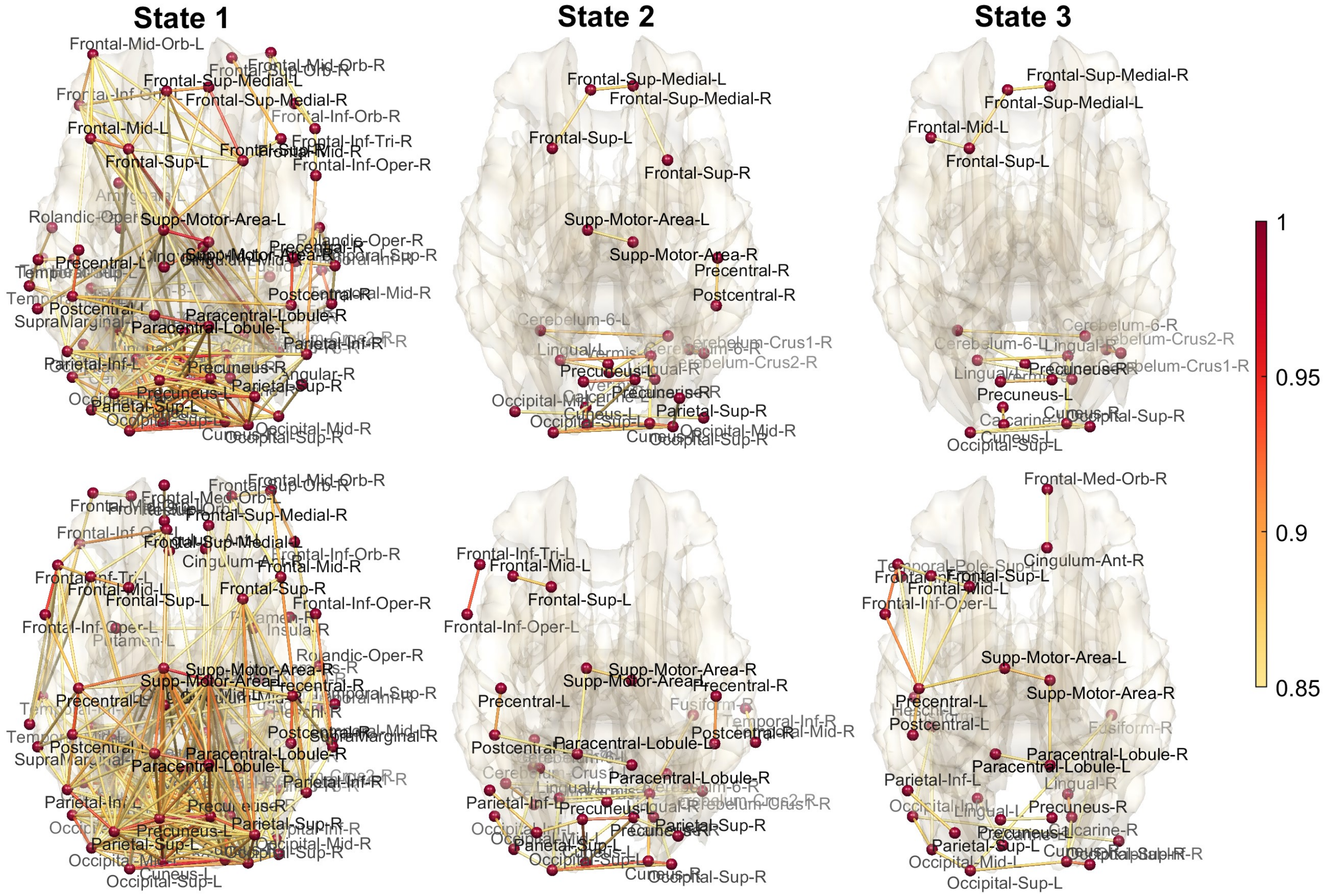}
\caption{The average correlation  within the estimated state of a single MZ-twin  based on the heat kernel method with FWHM 20 TRs. Within a state, correlations are averaged. Only connections above correlation 0.85 are displayed.}
\label{fig:twins_statespace}
\end{figure}

 We investigated if the state change pattern itself  is genetically heritable. Figure \ref{fig:statevisits} displays the state visits in randomly selected 3 MZ- and 3 DZ-twins. However, the time series of state changes do not synchronize between twins. Thus, we investigated the heritability of estimated state space. For each subject,  we computed the average correlation-map of each state, where the average is taken within each state.  Figure \ref{fig:twins_statespace} displays the average correlation map of a paired twin. The correlation at each connection is used as the input to the twin network analysis below.

\begin{figure}
\centering
\includegraphics[width=1\linewidth]{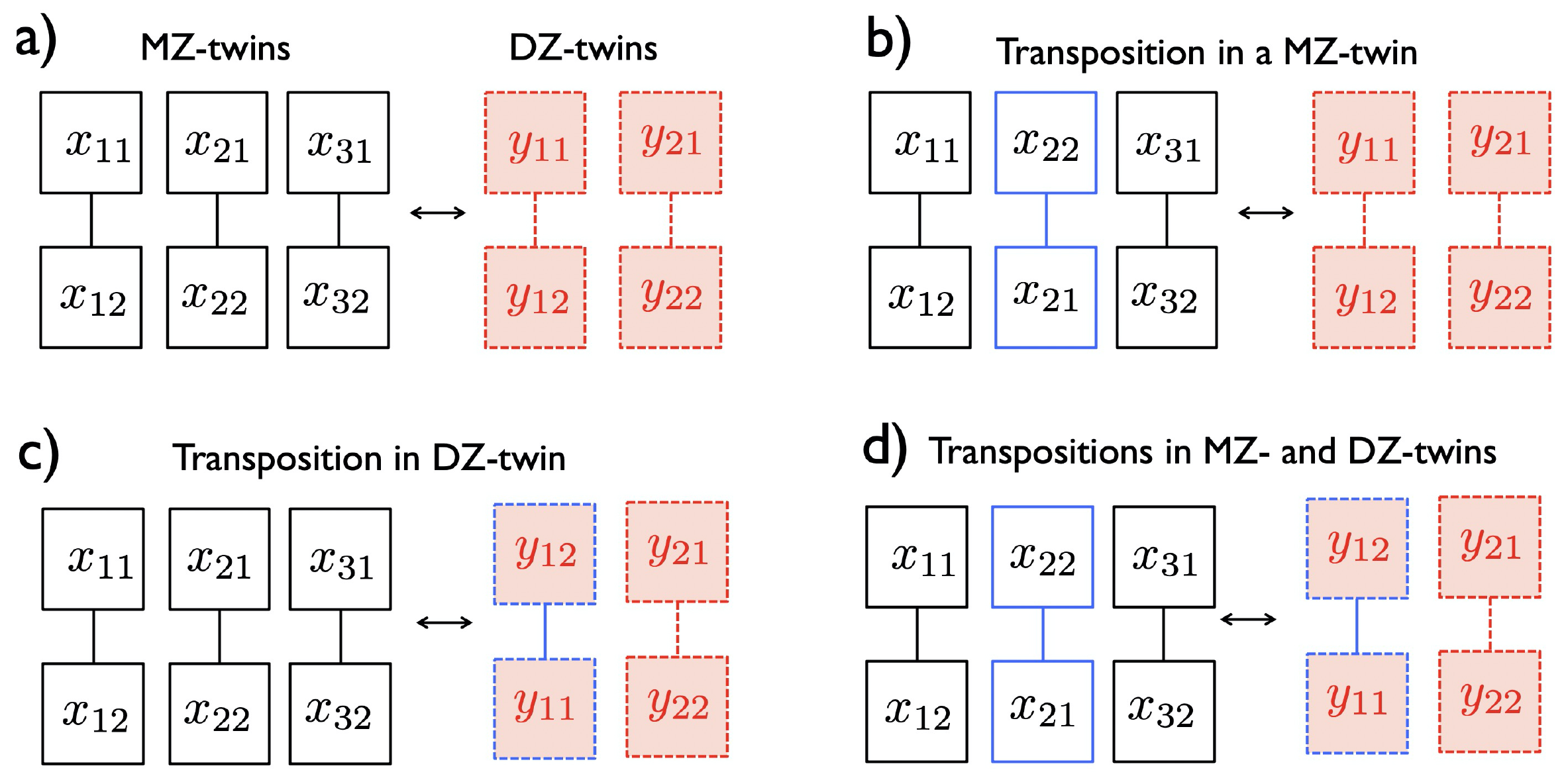}
\caption{The schematic of transpositions on 3 MZ- and 2 DZ-twins. b) Transposition  within a MZ-twin. c) Transposition within a DZ-twin. d) Transpositions in both MZ- and DZ-twins. Any transposition will affect the heritability estimate so it is necessary to account for as many transpositions as possible.}
\label{fig:schematic}
\end{figure}

We assume there are $m$ MZ- and $n$ DZ-twins. At edge $q$, let $x_{i} = (x_{i1}, x_{i2})^{\top}$ be the  $i$-th twin pair in MZ-twin and $y_{i} = (y_{i1}, y_{i2})^{\top}$ be the  $i$-th twin pair in DZ-twin. They are represented as
\bq {\bf x} &=&
\left(
\begin{array}{ccc}
 x_{11}, & \cdots   &, x_{m1}   \\
 x_{12}, &  \cdots &,   x_{m2} \\
\end{array}
\right)\\ 
{\bf y} &=& 
\left(
\begin{array}{ccc}
 y_{11}, & \cdots   &, y_{n1}   \\
 y_{12}, &  \cdots &,   y_{n2} \\
\end{array}
\right).
\eq
Let ${\bf x}_j$ be the $j$-th row of ${\bf x}$, i.e., ${\bf x}_j = (x_{1j}, x_{2j}, \cdots, x_{mj})$. Similarly let  ${\bf y}_j = (y_{1j}, y_{2j}, \cdots, y_{nj})$. Then MZ- and DZ-correlations  are computed as 
\bq \gamma^{MZ} ({\bf x}_1,{\bf x}_2) &=& corr({\bf x}_1,{\bf x}_2)\\
\gamma^{DZ} ({\bf y}_1,{\bf y}_2) &=& corr({\bf y}_1,{\bf y}_2).
\eq
In  the widely used ACE genetic model, the heritability index (HI) $h$, which  determines the amount of variation due to genetic influence in a population, is  estimated using Falconer's formula \cite{falconer.1995,chung.2019.NN,arbet.2020}. MZ-twins share 100\% of genes while same-sex DZ-twins share 50\% of genes on average. Thus, the additive genetic factor $A$ and the common environmental factor $C$  are related as 
\bq \gamma^{MZ} &=&A + C\\ 
\gamma^{DZ} &=& A/2 + C,\label{eq:HI} 
\eq

where $\gamma^{MZ}$ and $\gamma^{DZ}$ are correlations computed within MZ- and DZ-twins. Thus HI $h$, which measures the contribution of $A$, is given by 
$$h({\bf x}, {\bf y}) = 2 ( \gamma^{MZ} - \gamma^{DZ} ).$$
However, the order of twins is interchangeable, and we can {\em transpose} the $i$-th twin pair in MZ-twin such that
\bq \tau_{i}({\bf x}_1) &=& (x_{11} \cdots x_{i-1,1}, x_{i2}, x_{i+1,1} \cdots x_{m1}),\\
 \tau_i({\bf x}_2) &=& (x_{12} \cdots x_{i-1,2}, x_{i1}, x_{i+1,2} \cdots x_{m2})\eq
and obtain another twin correlation $\gamma^{MZ}(\tau_i({\bf x}_1), \tau_i({\bf x}_2))$ \cite{chung.2019.CNI}.  Ignoring symmetry,  there are $2^{m}$ possible combinations in ordering the twins, which forms a permutation group. The size of the permutation group grows exponentially large as the sample size increases.  Computing correlations over all permutations is not even computationally feasible for large $m$ beyond 100. Figure \ref{fig:schematic} illustrates many possible transpositions within twins. This has been the main weakness of the ACE model. Thus, we propose a fast online computational strategy for ACE. We propose to perform a sequence of random transpositions and compute the twin correlation at each transposition.  

Over transposition $\tau_i$, the correlation changes from $\gamma^{MZ}({\bf x}_1, {\bf x}_2)$ to $\gamma^{MZ}( \tau_i ({\bf x}_1), \tau_i ({\bf x}_2))$ incrementally. We will determine the exact increment over the transposition. The Pearson correlation between ${\bf x}_k$ and ${\bf x}_l$ involves the following functions. 
\bq \nu ({\bf x}_k) &=&  \sum_{l=1}^m x_{lk}\\
  \omega({\bf x}_k, {\bf x}_l) &=&  \sum_{r=1}^m \big(x_{rk}-\nu({\bf x}_k)/m \big) \big(x_{rl}-\nu({\bf x}_l)/m \big). \eq
The functions $\mu$ and $\omega$ are updated over transposition $\tau_i$ as
\bq \nu (\tau_i({\bf x}_k)) &=& \nu({\bf x}_k) - x_{ik} + x_{il} \\
\omega(\tau_i({\bf x}_k), \tau_i ({\bf x}_l))&= &\omega({\bf x}_k, {\bf x}_l)+ (x_{ik} - x_{il})^2/m\\
&&
 - (x_{ik} - x_{il}   )   \big(  \nu({\bf x}_k)  -\nu ({\bf x}_l) \big) /m.
\eq
Then the MZ-twin correlation after transposition is  updated as
$$
\gamma^{MZ}(\tau_i({\bf x}_1), \tau_i({\bf x}_2))= \frac{\omega(\tau_i({\bf x}_1), \tau_i({\bf x}_2))}{\sqrt{\omega(\tau_i({\bf x}_1), \tau_i({\bf x}_1)) \omega(\tau_i({\bf x}_2), \tau_i({\bf x}_2))}}. \label{eq:online-corr}
$$

The time complexity for correlation computation is 33 operations per transposition, which is substantially lower than the computational complexity of directly computing correlations per permutation. In the numerical implementation, we  sequentially apply random transpositions $\tau_{i_1}, \tau_{i_2}, \cdots, \tau_{i_J}$. This results in $J$ different twin correlations, which are averaged. Let $$\pi_1 = \tau_{i_1}, \pi_2 = \tau_{i_2} \circ \tau_{i_1}, \cdots, \pi_J = \tau_{i_J} \circ \cdots \circ  \tau_{i_2} \circ \tau_{i_1}.$$ 
The average correlation $\overline \gamma_J^{MZ}$ of all  $J$ transpositions is given by
$$ \overline \gamma_J^{MZ}= \frac{1}{J} \sum_{j=1}^J \gamma^{MZ} (\pi_{i_j}({\bf x}_1), \pi_{i_j}({\bf x}_2)).$$
In each sequential update, the average correlation can be updated iteratively as 
$$\overline \gamma_J^{MZ} =  \frac{J-1}{J} \overline \gamma_{J-1}^{MZ}  +  \frac{1}{J}\gamma^{MZ} (\pi_{i_J}({\bf x}_1), \pi_{i_J}({\bf x}_2)).$$
If we use enough number of transpositions, the average correlation $\overline \gamma_J^{MZ}$ converges to the true underlying twin correlation $\gamma^{MZ}$ for sufficiently large $J$. DZ-twin correlation $\gamma^{DZ}$ is estimated similarly and HI-map $h$ is given as the twice difference in twin correlations at each connection. 

\section{Applications}

\subsection{Dataset and preprocessing} 
The resting-state functional magnetic resonance images (rs-fMRI) were collected on a 3T
MRI scanner (Discovery MR750, General Electric Medical Systems, Milwaukee, WI, USA) with  a 32-channel RF head coil array. 
T1-weighted structural images (1 mm$^3$ voxels) were also acquired axially with an isotropic 3D Bravo
sequence (
TE = 3.2 ms, TR = 8.2 ms, TI = 450 ms, flip angle = 12$^\circ$). 
T2-weighted gradient-echo echo-planar pulse sequence images were collected during resting state with TE = 20 ms, TR = 2000 ms, and flip angle = 60$^\circ$.
The functional scans were undergone a series of data reduction, correction, registration,  and spatial and temporal preprocessing  \cite{burghy2016experience}.
The resulting rs-fMRI consists of  $91\times109\times91$  isotropic voxels at 295 time points.  
Excluding one subject that has no fMRI signals in two brain regions, the average fMRI signals of 479 healthy subjects 
ranging in age from 13 to 25 years were used. Among 479 subjects, there are 130 MZ-twins (59 male twins, 71 female twins) 
and 102 DZ-twins (51 male twins, 44 female twins, 7 opposite sex twins).

We employed the Automated Anatomical Labeling (AAL) brain template to parcellate the brain volume into 116 non-overlapping anatomical regions \cite{tzourio.2002}. The fMRI data were averaged across voxels within each brain region, resulting in 116 average fMRI signals with 295 time points for each subject. The rs-fMRI signals were then scaled to fit to unit interval [0, 1]. Then the data were made circular with circumference 2. For each subject, dynamically changing $116\times116$ correlation matrices in 116 regions were computed for all three methods. 


\subsection {Dynamically changing state space} 
\begin{figure}[t]
\centering
\includegraphics[width=1\linewidth]{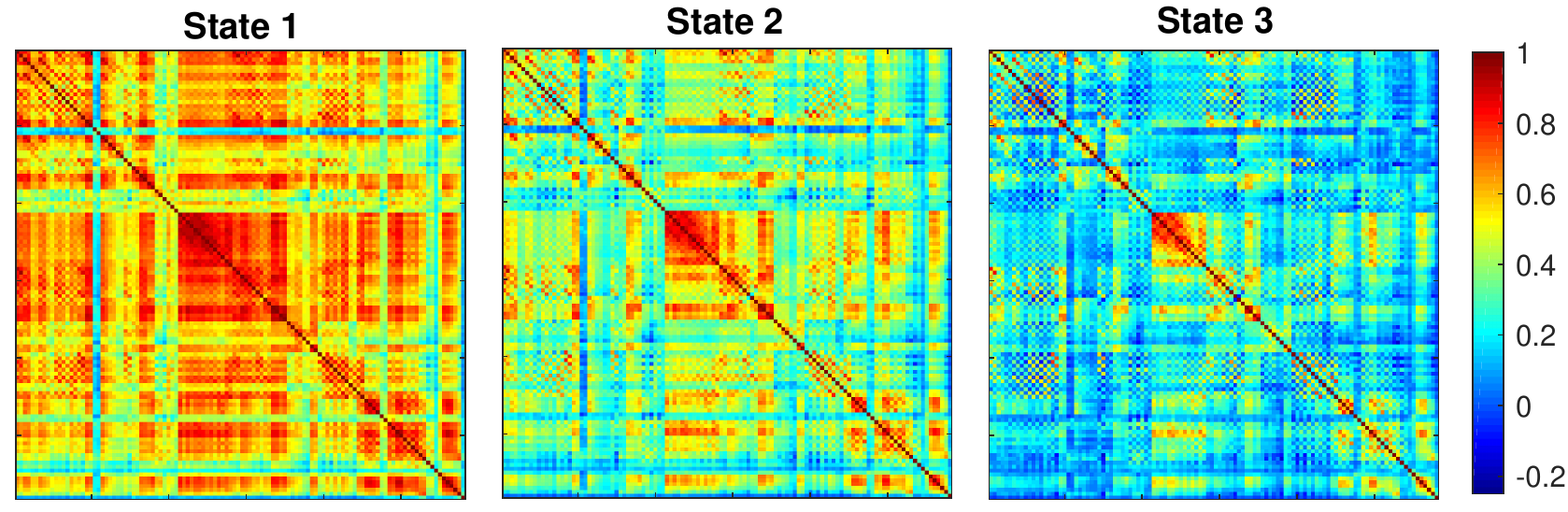}
\caption{The average correlation matrices (cluster centroids) of the three states.
The values of average correlations range from  0.03 to 0.98 in state 1, from -0.12 to 0.95 in state 2, and from -0.22 to 0.89 in state 3.
Only the windowless method with FWHM 20 TR was plotted because different methods have similar result.
Within each state, the absolute errors between the centroids obtained from different methods are all smaller than 0.075.}
\label{fig:AAL_Centroid}
\end{figure}

\begin{figure}[t]
\centering
\includegraphics[width=\linewidth]{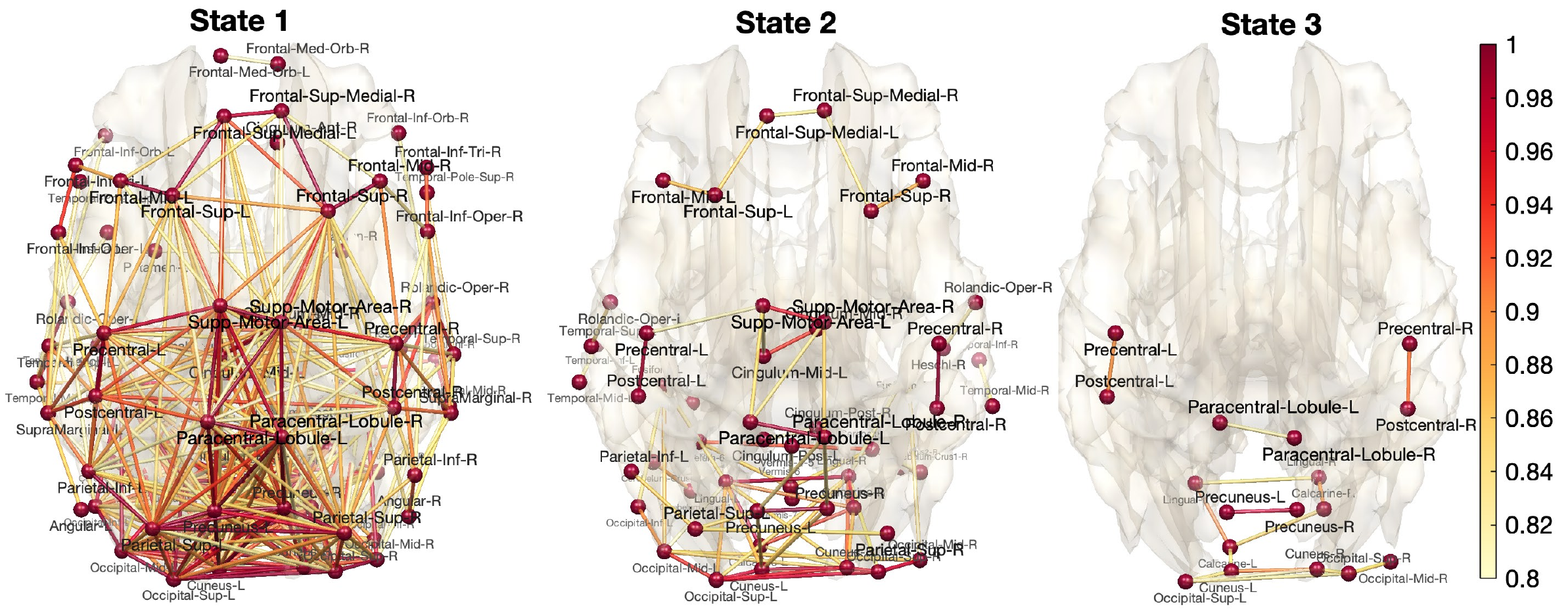}
\caption{The average correlation of the three states using the heat kernel method with FWHM 20 TRs.
Only strong connections with correlation above 0.8 are displayed.
State 1 includes all the strong connections in states 2 and 3. State 2 includes all the strong connections in state 3.
Left and right precunei, right superior parietal lobule, left cuneus and right lingual gyrus are the five most connected regions in 	state 1.}
\label{fig:AAL_Brain}
\end{figure}

\begin{figure}
\centering
\includegraphics[width=1\linewidth]{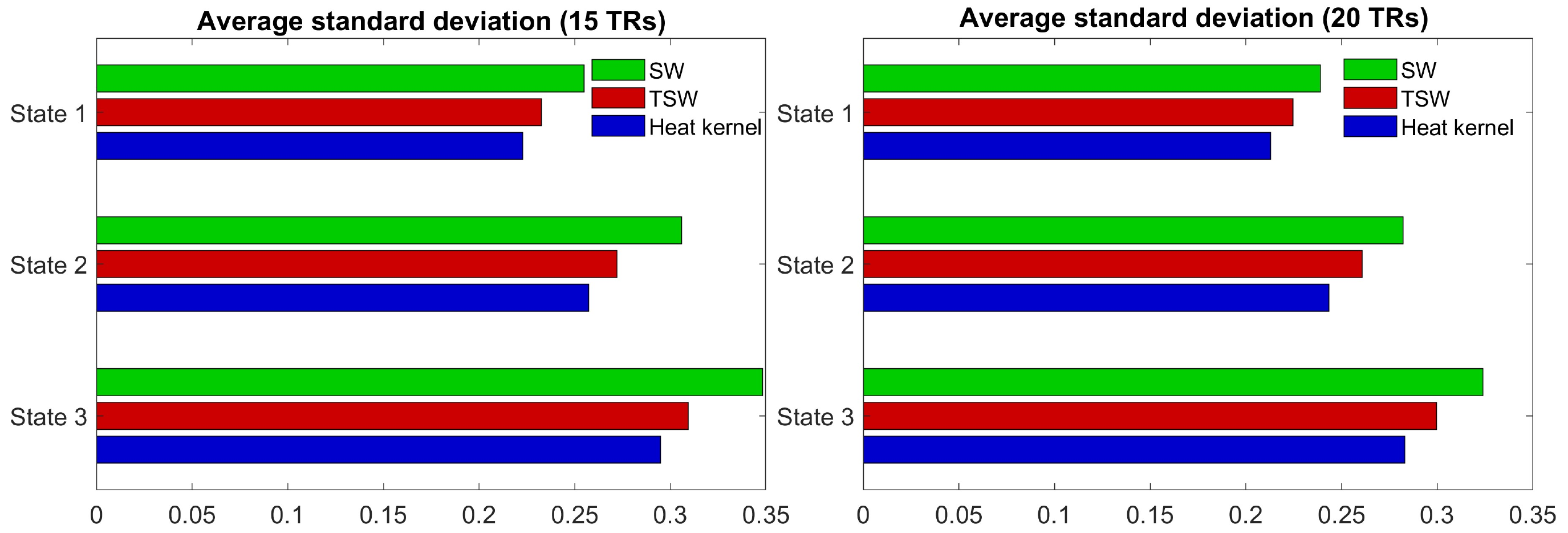}
\caption{The average standard deviation within each state. Heat kernel method had the smallest variability demonstrating the method estimates the state space more robustly. The state 1 has the strongest connectivity with the lowest amount of dispersion in connectivity.}
\label{fig:AAL_Std2}
\end{figure}

The dynamic correlation matrices for all the subjects were clustered into three states using the $k$-means clustering.
The average correlation within each state is displayed in Figures~\ref{fig:AAL_Centroid} and ~\ref{fig:AAL_Brain}. Within each state, we computed the standard deviation of correlations for each brain connection over all time points and subjects and then averaged them across all  brain connections (Figure \ref{fig:AAL_Std2}). For FWHM 15 TRs, relative to the SW-method, the average standard deviations within state 1, 2 and 3 were reduced  8.8\%,  11\% and   11.2\%  by the TSW-method, and reduced 12.5\%, 15.8\% and   15.3\%  by the heat kernel method, respectively. Similarly for FWHM 20 TRs, the average standard deviations were reduced 7.5\%,   7.6\%, and    5.9\%  by the TSW-method, and reduced 12.6\%,  13.8\%, and   10.8\% by the heat kernel method. The heat kernel method had the smallest variability within each state compared SW- and TSW-methods demonstrating the method estimates the state space more accurately.\\

\begin{figure}[t]
\centering
\includegraphics[width=\linewidth]{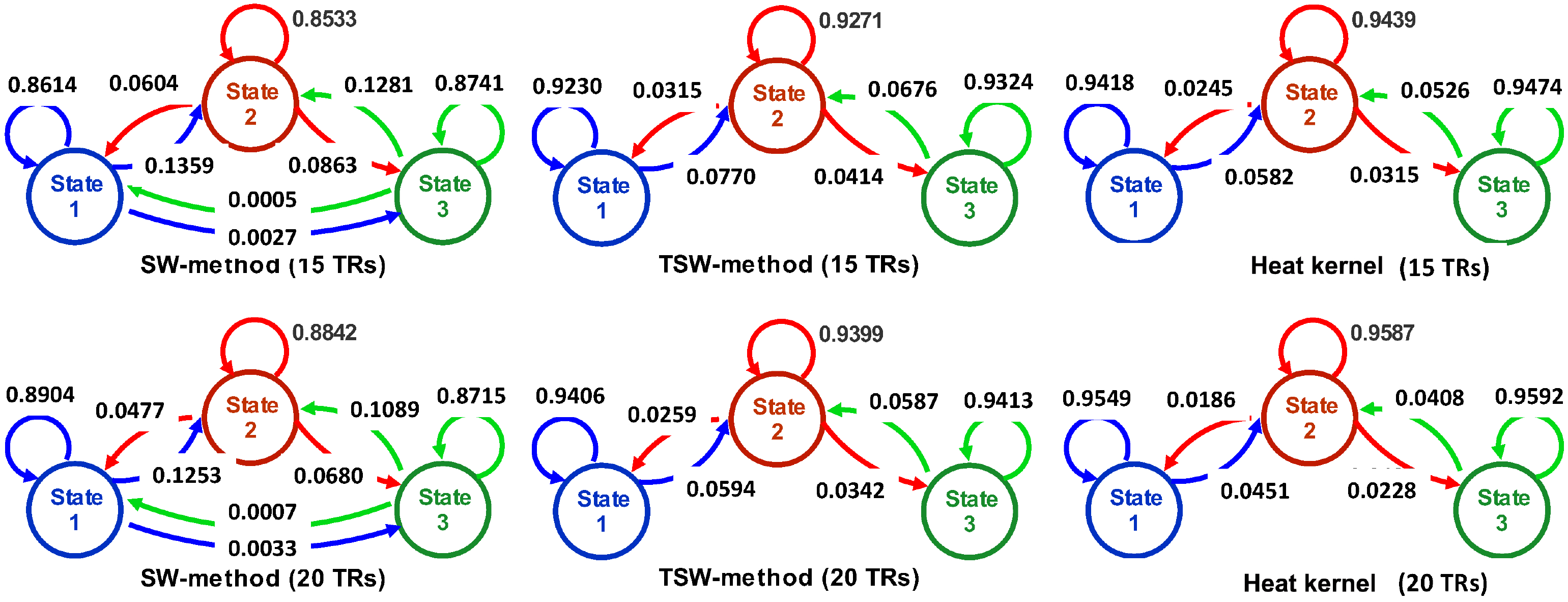}
\caption{Transition probabilities averaged across all 479 subjects for FWHM 15 TRs (top) and 20 TRs (bottom). 
In average, subjects remained in the same state for a long period of time before transiting to other states. 
The probabilities of remaining in the same state increase when  larger bandwidth is used.}
\label{fig:AAL_Transition}
\end{figure}

\begin{figure}
\centering
\includegraphics[width=1\linewidth]{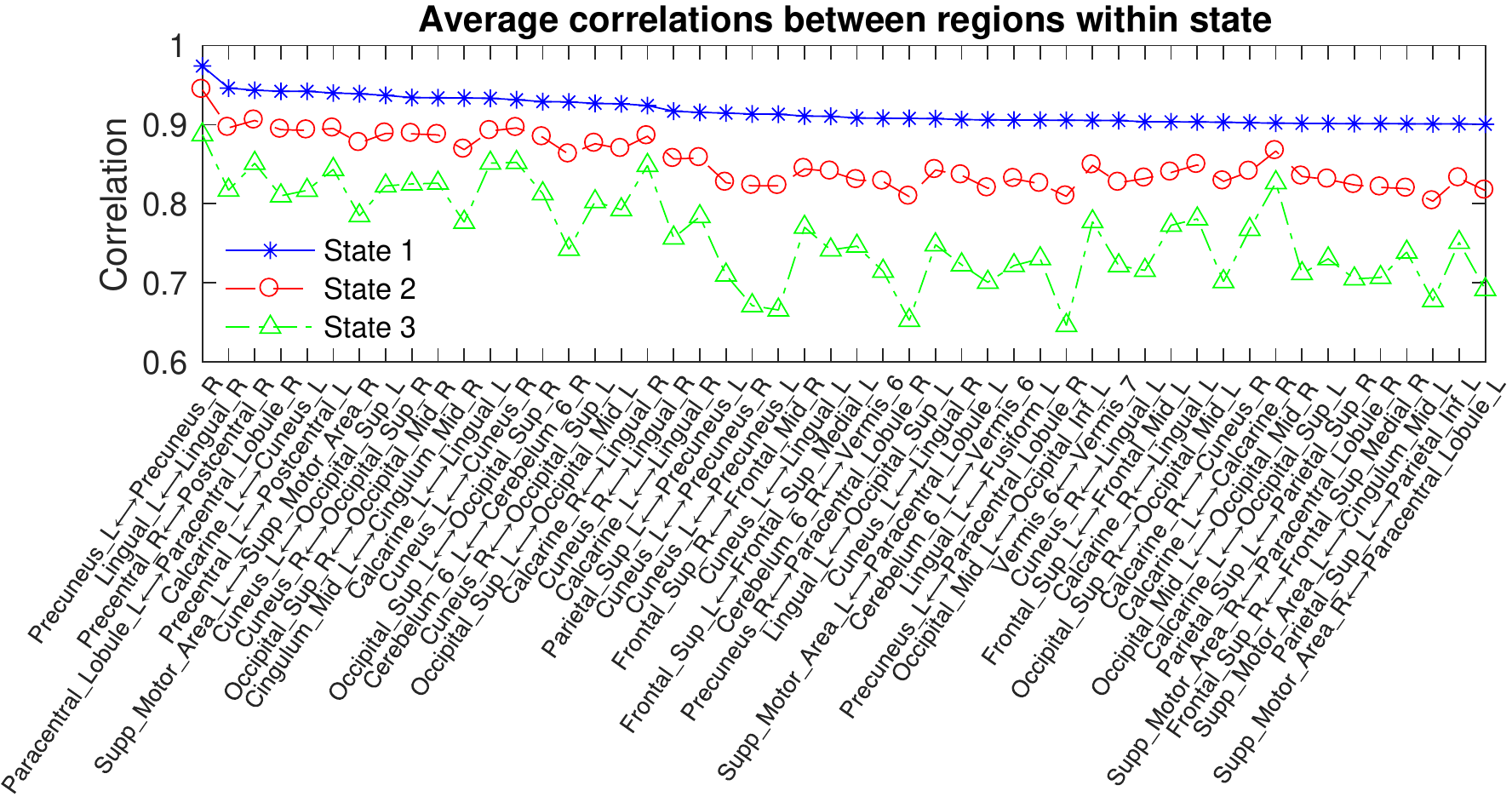}
\caption{50 strongest connections in state 1. They are sorted according to correlation values. 11 of these connections are  interhemispheric connectivity demonstrating high synchronization across hemispheres.}
\label{fig:connection_centroid}
\end{figure}

{\em Transition probability between states.} For the proposed heat kernel method, we computed the transition probability between states and averaged over all subjects (Figure \ref{fig:AAL_Transition}).
In average, each subject remained in the same state for a long period of time before transitioning to other states. The average probability of staying in the same state at any time point is between 0.85 and 0.94 for larger bandwidth (20 TRs) and between 0.88 and 0.96 for smaller bandwidth (15 TRs). Almost zero transition probability between state 1 and state 3 show the inability of transitioning directly between  these two states. The proposed heat kernel method has the lowest transition probabilities between different states and the highest probabilities of remaining in the same state demonstrating more stable state space estimation.\\

{\em Strong connections in state space.} 
From the average correlation of the three states (Figures \ref{fig:AAL_Centroid} and \ref{fig:AAL_Brain}), we tabulated the first 50 connections with  highest average correlations in state 1 (Figure \ref{fig:connection_centroid}).
Among those 50 connections, 11 are  hemispherically paired demonstrating strong hemispheric synchronization. Among these 11 paired regions, calcarine sulci, cunei, lingual gyri, superior occipital gyri and  middle occipital gyri also have strong connections among them. Figure~\ref{fig:AAL_Brain}-left displays strong connections with correlation values larger than 0.8 in state 1.

\subsection {Dynamic interhemispheric connectivity}

Because evidence shows that functional connectivity in state 1 is highly synchronized across hemispheres, we  decided to perform the separate  interhemispheric connectivity analysis. Excluding the 8 vermis regions that do not belong to the left or right brain hemisphere, we computed the 54 dynamic correlations of the 54 hemispherically paired brain regions using the proposed heat kernel method. For each of the 54 interhemispheric pairs, the dynamic correlations at 295 time points were concatenated across 479 subjects, which resulted in $295\cdot 479=141305$ total number of correlations that served as the input to $k$-means clustering.

\begin{figure}[t]
\centering
\includegraphics[width=1\linewidth]{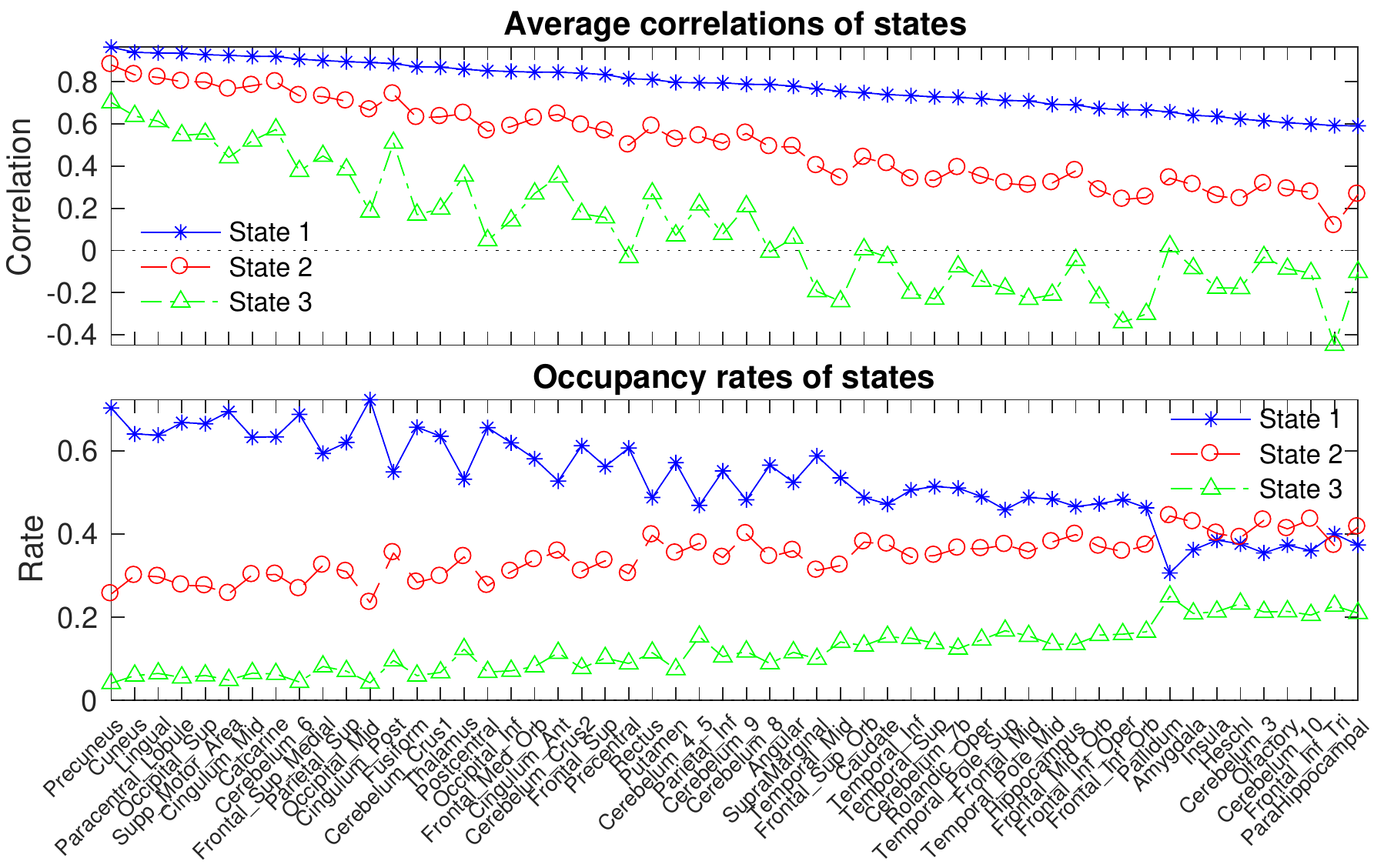}
\caption{Dynamic interhemispheric  correlations. They are sorted in terms of the average correlations of state 1.}
\label{fig:interhemisphere_centroid}
\end{figure}

The heat kernel method reduced rapid state changes and high-frequency fluctuations caused by the use of discrete windows. Figure \ref{fig:interhemisphere_centroid} displays the  average correlation  and the occupancy rate  of each state and interhemispheric  connectivity \cite{yaesoubi2015dynamic,ombao2018statistical}. 
Precuneus,  cuneus, lingual gyrus, paracentral lobule and superior occipital are the five brain regions having the highest interhemispheric correlations in the state space, and thus have the strongest hemispheric symmetry. The parahippocampal gyrus, inferior frontal gyrus (pars triangularis),  lobule X of cerebellar hemisphere, olfactory cortex and lobule III of cerebellar hemisphere are the five brain regions having the weakest hemispheric symmetry.

\subsection {Heritability of state space}

\begin{figure}[t]
\centering
\includegraphics[width=1\linewidth]{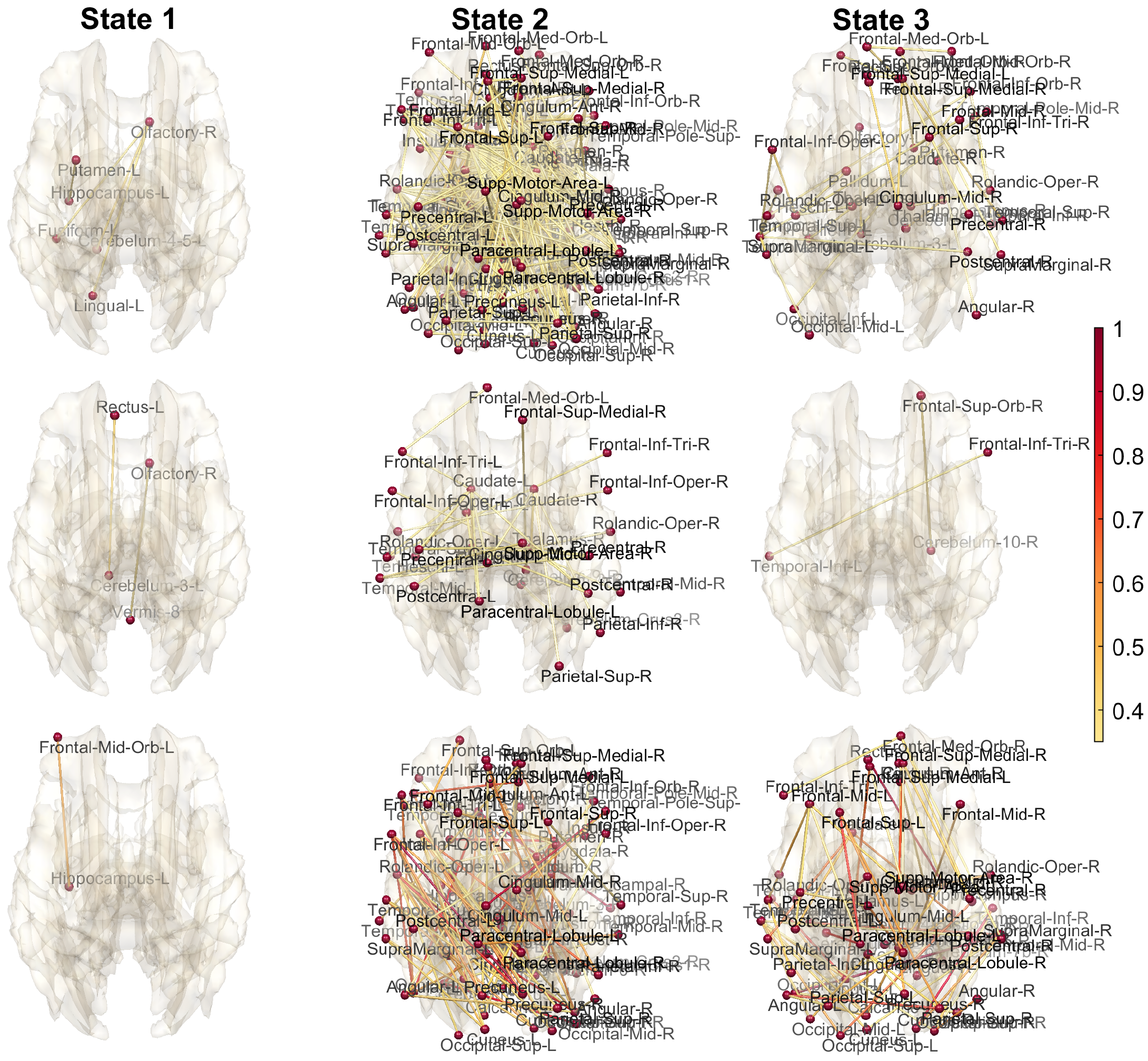}
\caption{Twin correlation of state space. The 1st and 2nd rows display MZ- and DZ-correlation of each state above 0.35. The 3rd row displays the HI of each state. The twin correlations and HI are estimated through the  transposition method. Only connections with HI above 0.75 are shown.}
\label{fig:twins_HI}
\end{figure}

We applied the proposed transposition method to the time series of state visits and computed the heritability. We further investigated the genetic contribution of the dynamic changes of the three distinct states based on 130 MZ- and 102 DZ-twins.  It is unclear the degree of heritability of such states. The heritability of these states has never been investigated before. Subsequently, the heritability of the estimated state space is computed using the proposed transposition method. 

We randomly transpose a twin and update the correlation using the proposed online formula 50000 times. This process is repeated 100 times and total $50000 \times 100$ correlations are averaged to obtain the underlying MZ-twin correlation. Similarly we estimated the DZ-twin correlation. At each connection, the standard deviation of 100 results was smaller than 0.01, which guarantees the convergence of the estimate within two decimal places in average. Figure \ref{fig:twins_HI} displays the twin correlations and the estimated HI-map of each state. The standard derivation (SD) of the estimated HI is bounded by the twice the sum of SD of twin correlations, i.e.,
$$ SD (h) \leq 2 [SD (\gamma^{MZ}) +SD (\gamma^{DZ})].$$
Thus, the SD of HI is within 0.04. The HI-map is the baseline index and may be interpreted as determining the percentage genetic contributions. 
 
 \begin{table}
\centering
\resizebox{\columnwidth}{!}{
    \begin{tabular}{ccc}
    \hline
    \multicolumn{1}{l}{State} & Connection & HI \\ \hline
    \multirow{5}{*}{1} 
    & Left hippocampus \;-\; Left middle frontal gyrus (orbital) & 0.82 $\pm$ 0.04\\
    & Right caudate nucleus \;-\; Left middle frontal gyrus (orbital) & 0.72 $\pm$ 0.04\\
    & Left hippocampus \;-\; Left middle frontal gyrus (lateral) & 0.70 $\pm$ 0.04\\
    & Lobule IV, V of vermis \;-\; Left middle frontal gyrus (orbital) & 0.69 $\pm$ 0.04\\
    & Left supramarginal gyrus \;-\; Left hippocampus & 0.68 $\pm$ 0.04\\ \hline
    \multirow{5}{*}{2} 
    & Right precuneus \;-\; Left opercular part of inferior frontal gyrus & 0.99 $\pm$ 0.04\\
    & Right inferior temporal gyrus \;-\; Left olfactory cortex & 0.98 $\pm$ 0.04\\ 
    & Right superior temporal pole \;-\; Left middle cingulate & 0.98 $\pm$ 0.04 \\
    & Left lobule IV, V of cerebellar hemisphere \;-\; Left rolandic operculum & 0.96 $\pm$ 0.04\\
    & Right amygdala \;-\; Left amygdala & 0.95 $\pm$ 0.04\\ \hline 
    \multirow{5}{*}{3} 
    & Right middle temporal gyrus \;-\; Left inferior occipital & 1.00 $\pm$ 0.04\\
    & Right hippocampus \;-\; Left gyrus rectus & 0.99 $\pm$ 0.04\\
    & Right inferior temporal gyrus \;-\; Right cuneus & 0.96 $\pm$ 0.04\\
    & Left inferior temporal gyrus \;-\; Left lingual gyrus & 0.93 $\pm$ 0.04\\
    & Left lobule VIII of cerebellar hemisphere \;-\; Left paracentral lobule & 0.93 $\pm$ 0.04\\ \hline \\
    \end{tabular}
}
\caption{Top five most heritable connections in each state.}
\label{tab:HItable}
\end{table}

The highly heritable connections show very different network pattern than the group-level average state space estimation. Strong connections are not necessarily heritable. In fact, many connections with low correlations in states 2 and 3 are most heritable. Table \ref{tab:HItable} lists 5 most heritable connections in each state. In state 1, left hippocampus and left middle frontal gyrus (orbital) connection shows the strongest heritability among many other connections. In state 2, right precuneus and left opercular part of inferior frontal gyrus connection shows the strongest heritability. In state 3, right middle temporal gyrus and left inferior occipital connection shows  the strongest heritability.

\section{Discussion}

The resting-state networks tend to remain in the same state for a long period before the  transition to another state \cite{Allen2014,Shakil2016,calhoun2016time,abrol2017replicability,nielsen2018predictive}. In this study, the proposed heat kernel method showed a longer stability with less rapid changes in the state space and exhibited a higher probability of remaining in the same state compared to the SW- and TSW-methods. We have further shown  that the proposed heat kernel method is robust over different choice of bandwidth  (15 and 20 TRs). 

In this study, the average correlation matrices of the three states follow similar connectivity patterns to the previous studies \cite{haimovici2017wakefulness,cai2018estimation}. We observed  higher correlations between calcarine sulci, cunei, lingual gyri, superior occipital gyri and  middle occipital gyri. All these regions belong to the occipital lobe and the part of visual network that are often observed in the resting state networks. Compared to other resting state networks, the  visual network has the strongest connectivity across different states, followed by the somatomotor network  \cite{ting2018multi,al2019tensor}.

Previous {\em static} network studies have demonstrated high correlations between hemispherically paired brain regions \cite{salvador2005neurophysiological,stark2008regional,zuo2010growing,anderson2010decreased}.
In \cite{salvador2005neurophysiological}, it was shown that the median cingulate and paracingulate gyri, thalamus, precuneus, anterior cingulate and paracingulate gyri are some of the regions with highest interhemispheric correlations.
\cite{stark2008regional} demonstrated higher interhemispheric correlation in primary sensory-motor cortices,
including postcentral gyrus, occipital pole, lingual gyrus, cuneal cortex, precentral gyrus among other regions.
In \cite{anderson2010decreased}, the authors showed a trend toward higher interhemispheric connectivity near the midline, such as the frontal pole, occipital cortex and medial parietal lobe, deep gray nuclei, and cerebellum. 

In this paper, we demonstrated that the {\em dynamic} change of brain network is also highly symmetric across hemispheres. The results showed that hemispherically paired regions with high correlations in state 1 also have high correlations in state 2. Further, states 1 and 2 dominate the dynamic network changes with the occupancy rate over 75\%. Consistent with previous statistic network studies,  we observed relatively higher interhemispheric correlations in precuneus, cuneus, lingual gyrus, paracentral lobule, superior occipital, supplementary motor area, midcingulate area and calcarine sulcus among other regions. Many of these regions are close to the midline as well.

We also investigated the heritability of the estimated state spaces and the pattern of state changes. 
The dominant connections in each state are not necessarily heritable. We observed states 2 and 3 have far more connections with high heritability than state 1. We developed the novel transposition method that  speed up generating permutations for accurate estimation of heritability. The transposition methods can be easily adapted a faster alternative to the permutation test.

\section*{Acknowledgements}

We would like to thank Chee-Ming Ting and Hernando Ombao of  KAUST and Martin Lindquist of Johns Hopkins University for providing valuable discussions  on the state space modeling. We would like to thank Yixian Wang of University of Wisconsin-Madison, Andrey Gritsenko of Northeastern University and Li Shen of University of Pennsylvania for providing valuable discussion on twin correlations.

\bibliographystyle{IEEEtran}
\bibliography{reference.2020.07.01,SGHuangWisc.TBME.2019.05.12}

\end{document}